\documentclass{article}

\usepackage{icrc2011}

\title{{\it \bf Fermi} Gamma-Ray Burst Monitor Science Highlights}

\shorttitle{Connaughton {\it Fermi} GBM Highlights}

\authors{V. Connaughton$^{1,2}$}
\afiliations{$^1$CSPAR and Department of Physics, University of Alabama in Huntsville, Huntsville, AL 35805\\
$^2$for the GBM Team}
\email{valerie@nasa.gov}

\abstract{Three years after the launch of the {\it Fermi} Gamma-ray Space Telescope, both
of its scientific instruments are operating perfectly and continuing to make breakthroughs in
astrophysics, particle physics, and atmospheric science.  I report here on the
highlights of the scientific program of the {\it Fermi} Gamma-ray Burst Monitor (GBM).}

\begin{document}
\maketitle

\section{The GBM instrument}

The Gamma-ray Burst Monitor (GBM) is the secondary instrument on board the {\it Fermi}
Gamma-ray Space Telescope, and is the successor to the Burst And Transient Source
Experiment (BATSE) on the Compton Gamma-ray Observatory.
Heritage from BATSE is evident in GBM detector technology, scientific agenda,
and in the composition of the team which designed, built and contributes to its operation.
The main science objective of GBM is to complement the observations of Gamma-Ray Bursts (GRBs)
by the primary instrument on board {\it Fermi}, the Large Area Telescope (LAT) \cite{atwood}.

GBM is composed of 14 scintillators, distributed around the spacecraft to provide a full view of the
sky, with lesser sensitivity to regions far off the z-axis
axis of the spacecraft, which is aligned with the LAT
boresight.  Through the use of two different types of
scintillating materials, GBM has broader energy coverage than its predecessor, BATSE.  At low energies, 8 keV - 1 MeV,
thin (1.27 cm) Sodium Iodide (NaI) disks with differing orientations provide a directional response that enables localization
of observed sources to accuracies of a few degrees on the sky \cite{vc:loc}.  They are complemented for studies of energy spectra
by two thicker (12.7 cm) Bismuth Germanate (BGO), which are sensitive between 200 keV and 40 MeV.  All 14
detectors are powered by a single Power Supply Box (PSB), with independent high-voltage supplies to the photomultiplier tubes
(PMTs) encased with each scintillator.  A Data Processing Unit provides control of the PSB, readout and digitization of
the detector signal, and communication with buses on the spacecraft that allow commanding of the instrument and packaging
of the scientific and diagnostic data for transfer to the ground.  A full description of GBM is given in \cite{meegan09}.

 \begin{figure*}
  \vspace{5mm}
  \centering
  \includegraphics[width=6.in]{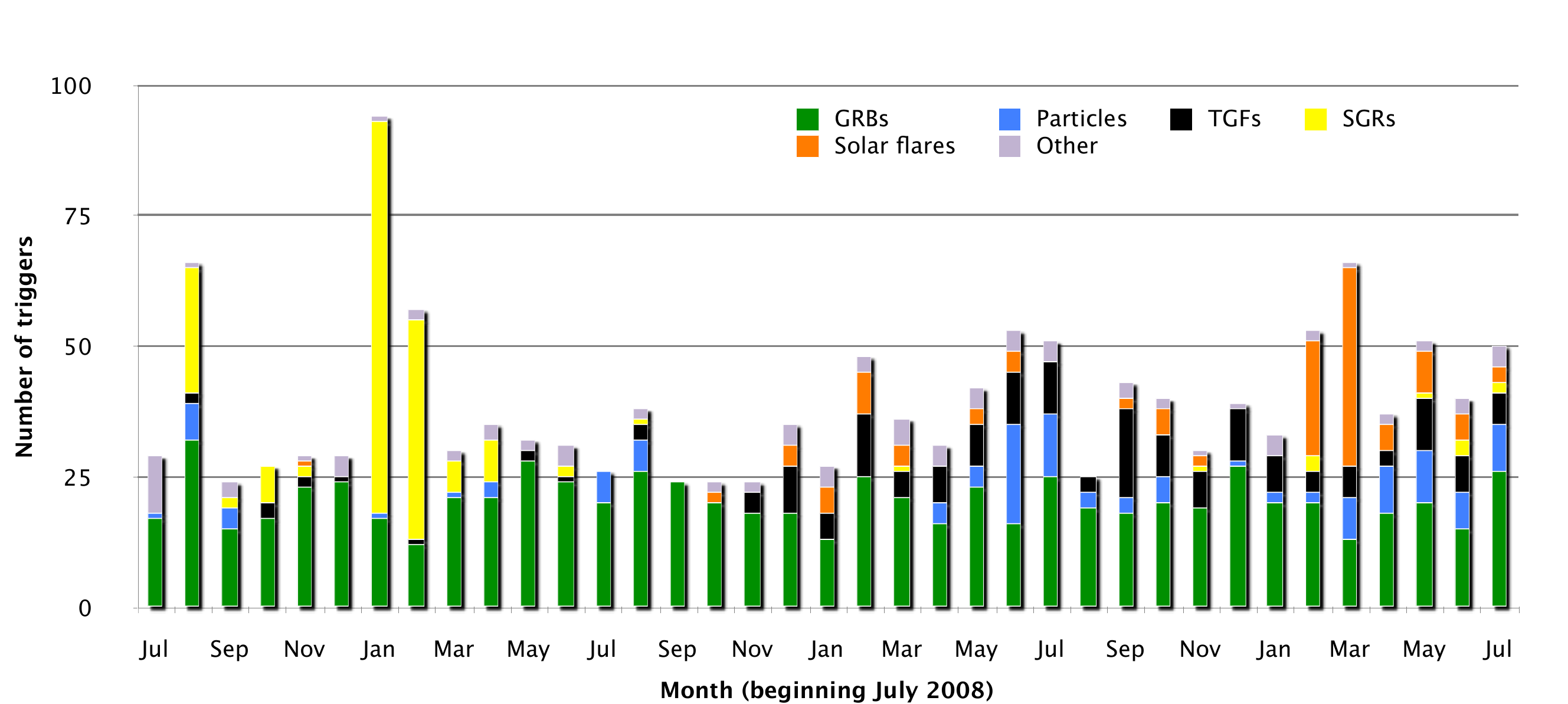}
  \caption{Monthly trigger rate for GBM over the first three years of operation.}
  \label{trigs}
 \end{figure*}

\section{GBM Operation}

GBM records data continuously throughout its 90 minute orbit, except during passages through the
South Atlantic Anomaly (SAA)
where the high-voltages to the PMTs are disabled to prevent life-shortening damage to the PMTs.
The unshielded scintillators record a mixture of cosmic-ray background (primaries and secondaries)
and contributions from gamma-ray sources (mostly galactic).  Cosmic-ray background
contributions dominate, with event rates that vary according to geomagnetic latitude, but are about
one [two] thousand counts per second for the NaI [BGO] detectors.  The
event rate variation and relative contributions are energy-dependent, and may also be affected by solar
and magnetospheric activity.  GBM is thus a heavily background-limited instrument and is most effective at detecting a source
which varies on time-scales that can be distinguished from the orbital variations of the background.  This is accomplished in three
ways: triggering on impulsive sources, measurements of count rate changes when sources rise from and set behind the Earth,
and extraction of periodic signals indicating pulsar activity from the residuals
in the data after careful cleaning, fitting, and subtraction of
measured trends in detector count rates.

Triggering on impulsive sources occurs when at least two detectors register a statistically significant increase
in count rates above background (16 s average)
in one or more time intervals between 16 and 4096 ms in at least one of four energy bands
(nominally 25-100 keV, 50-300 keV, above 100 keV, or above 300 keV).  Currently, 30 trigger algorithms are deployed, with
overlapping time windows and significance levels designed to
 maximize the probability of detecting sources of interest while minimizing nuisance triggers from magnetospheric activity
or accidental triggers.  All but four of the algorithms operate solely using the NaI detectors, with algorithms using the BGO
detectors
implemented after launch to maximize sensitivity to Terrestrial Gamma-ray Flashes (TGFs).  The 25-100 keV algorithms are
now restricted to short time-scales (less than 0.256 s) to reduce the number of particle triggers while keeping sensitivity to the
main source of low-energy triggers, Soft Gamma-ray Repeaters (SGRs).
The 50-300 keV energy range is the canonical BATSE trigger band, which
maximizes sensitivity to GRBs and covers the energy range where background rates are most stable.  Figure \ref{trigs} shows
the 3-year trigger history of GBM, with the color-coding reflecting the nature of the trigger.  The deployment and removal
of algorithms can be discerned in the decrease of noise triggers and increase in TGF triggers.  Increasing solar activity is
reflected in the solar flare trigger rate, whilst the sporadic activity of SGR sources follows a less predictable pattern.

GBM in trigger mode is not especially sensitive to most galactic sources, which are either relatively steady
 or vary on time-scales longer than 4 s.  Because their emission is typically more persistent than seen in triggered
bursts,
GBM can accumulate a source signal over long exposures by measuring and accumulating the slight steps in count rates
in detectors which have a favorable viewing angle when the source sets behind and rises from the Earth's limb.  This
Earth Occultation technique requires careful fitting of background rates in the time window before and after
the expected
step, including steps from interfering sources which may be occulted in the same time window, and uses the modeling
of the time-varying response of the GBM detectors during the exposure.  In this way, source fluxes can be
calculated over a long baseline, with about a 100 mCrab daily sensitivity at 20 keV,
that increases to just under 10 mCrab over long exposures
taking into account systematic effects. The Earth Occultation technique is described in
\cite{cwh:catalog}, with
two catalogs, one above 100 keV \cite{case} and one covering observations of 209 monitored sources above 8 keV,
of which 109 sources are high-confidence detections \cite{cwh:catalog}.
Long-term lightcurves for all the monitored sources are regularly updated and available online\footnote{
http://heastro.phys.lsu.edu/gbm}.

Many galactic sources of hard X-rays also exhibit periodic signals which can be
extracted by fitting and subtracting Fermi's orbital variations and source occultation steps in the
continous GBM data, excluding regions of impulsive activity, and searching for periodicity from known and unknown sources in
the residuals.  Using this pulsar technique, GBM is sensitive to periodic sources with frequencies between 1 mHz and
2 Hz,  and monitors 32 accretion-powered pulsars, of which 26 are detected, 18 in outburst.
Long-term pulsed flux lightcurves and frequencies are regularly updated and available online\footnote{
http://gammaray.nsstc.nasa.gov/gbm/science/pulsars}.

These three methods, triggering, Earth Occultation, and pulsar techniques, allow GBM to pursue
a broad scientific agenda, some highlights of which are presented in the following
sections on GRBs, galactic sources, and solar system science.

\section{Gamma-Ray Bursts}

 \begin{figure*}
  \vspace{5mm}
  \centering
  \includegraphics[width=5.5in]{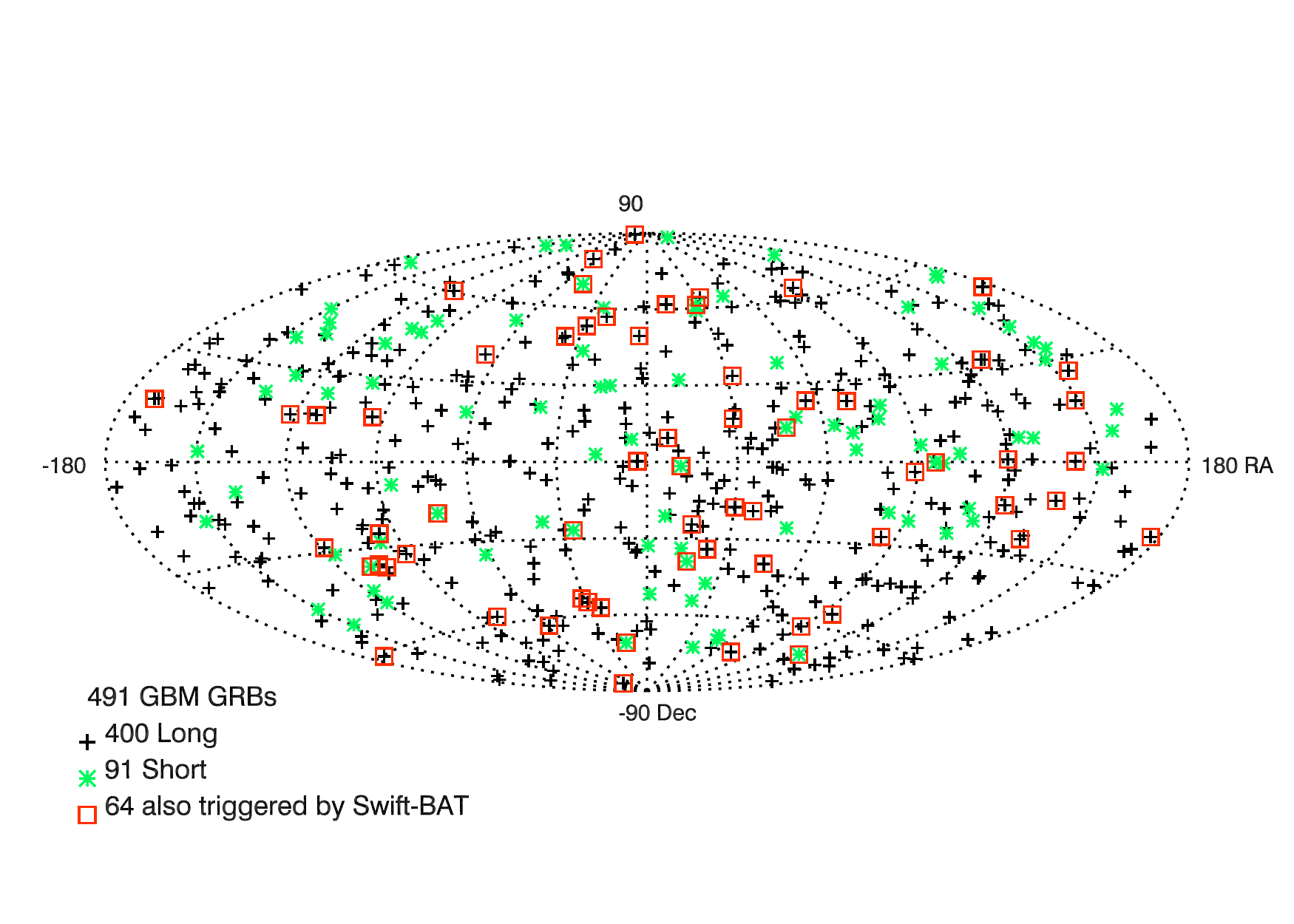}
  \caption{Distribution of GRB coordinates in celestial coordinates for the GRBs detected
by GBM in the first two years of operation.  Short bursts are light stars, long bursts pluses.
GBM-triggered GRBs which also triggered the Swift-BAT are indicated by open squares.}
  \label{skyfig}
 \end{figure*}

The primary scientific mission of GBM is the observation of GRBs.
In combination with the LAT, GBM enables broad-band spectroscopy over nearly seven decades of energy
of the GRBs detected in common.  By itself, GBM triggers on approximately 250 GRBs per year, providing
localizations, lightcurves and a measure of the intensity of the GRB in near real-time through the
GRB Coordinates Network\footnote{
http://gcn.gsfc.nasa.gov/} and at the Fermi Science Support Center (FSSC). The operations team delivers to the FSSC full
science data, following regular downlink, hours after the burst, and
processed data, including instrument responses and refined localizations, in the days that follow.
GBM is a prolific burst detector because of its 85\% duty cycle (outside SAA passages) and wide field-of-view, so that
any point in the sky has about a 50\% coverage (with high declination sources visible all the non-SAA time
during certain times in the {\it Fermi} precession cycle). Although the localization capabilities of GBM are modest
compared to imaging instruments, they are good enough to assist searches
for electromagnetic counterparts to gravitational wave candidates, and
for plausible GRBs that might be related to central-engine-driven supernovae.
Two such searches \cite{soderberg, corsi} revealed no GRB candidates that might be associated with the relativistic outflows
from Type 1bc and 1c supernovae.

A sky-map of the 491 GRBs from the first GBM GRB catalog \cite{wsp:cat}, which covers two years of observations,
 can be seen in Figure \ref{skyfig}, showing that
the GRBs are distributed isotropically on the sky and that the fraction of short-to-long bursts (20\%) is
comparable to that seen by BATSE.
Spectral analyses of these GRBs
are presented in \cite{ag:cat} and a searchable database of the catalog is available through the FSSC\footnote{
http://heasarc.gsfc.nasa.gov/W3Browse/fermi/fermigbrst.html}.

The BGO detectors provide enhanced high-energy coverage relative to BATSE, allowing
analyses with good statistics
of the MeV emission from GRBs \cite{eb}, and enabling detailed studies of the spectral \cite{guiriec:short}
and temporal \cite{bhat:pulses} characteristics of the brightest short bursts, which generally have spectra that turn over
at higher energies than long bursts.  Comparing short and long bursts, Guiriec et al. \cite{guiriec:short} find
that short bursts are similar to long bursts in their time-resolved
spectral characteristics, but with behavior that is stretched to
higher energies. Ghirlanda, Ghisellini, \& Nava \cite{ghirlanda} conclude from their analysis of GBM spectral data
that the emission mechanism in short and long bursts may be very similar.

 \begin{figure}
  \vspace{5mm}
  \centering
  \includegraphics[width=3.in]{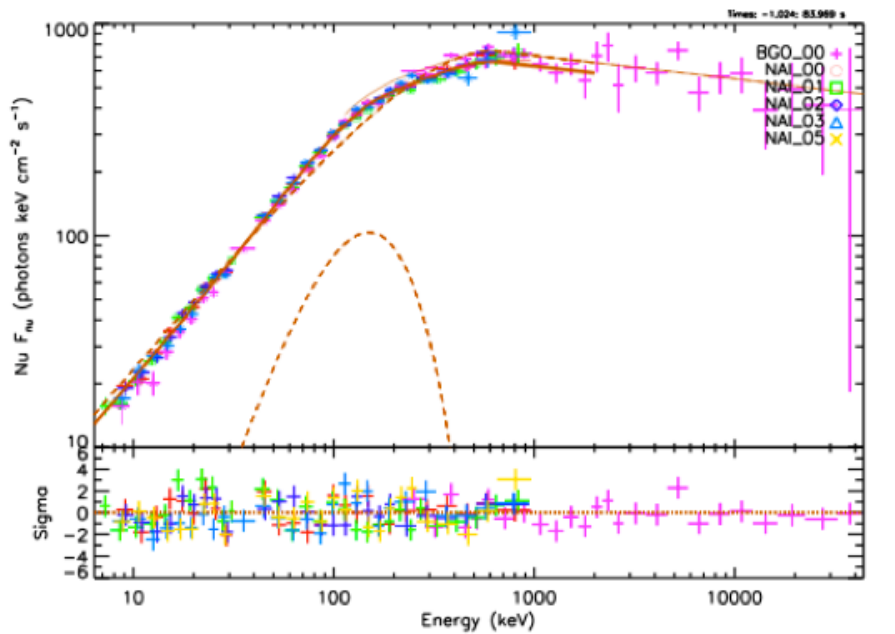}
  \includegraphics[width=3.in]{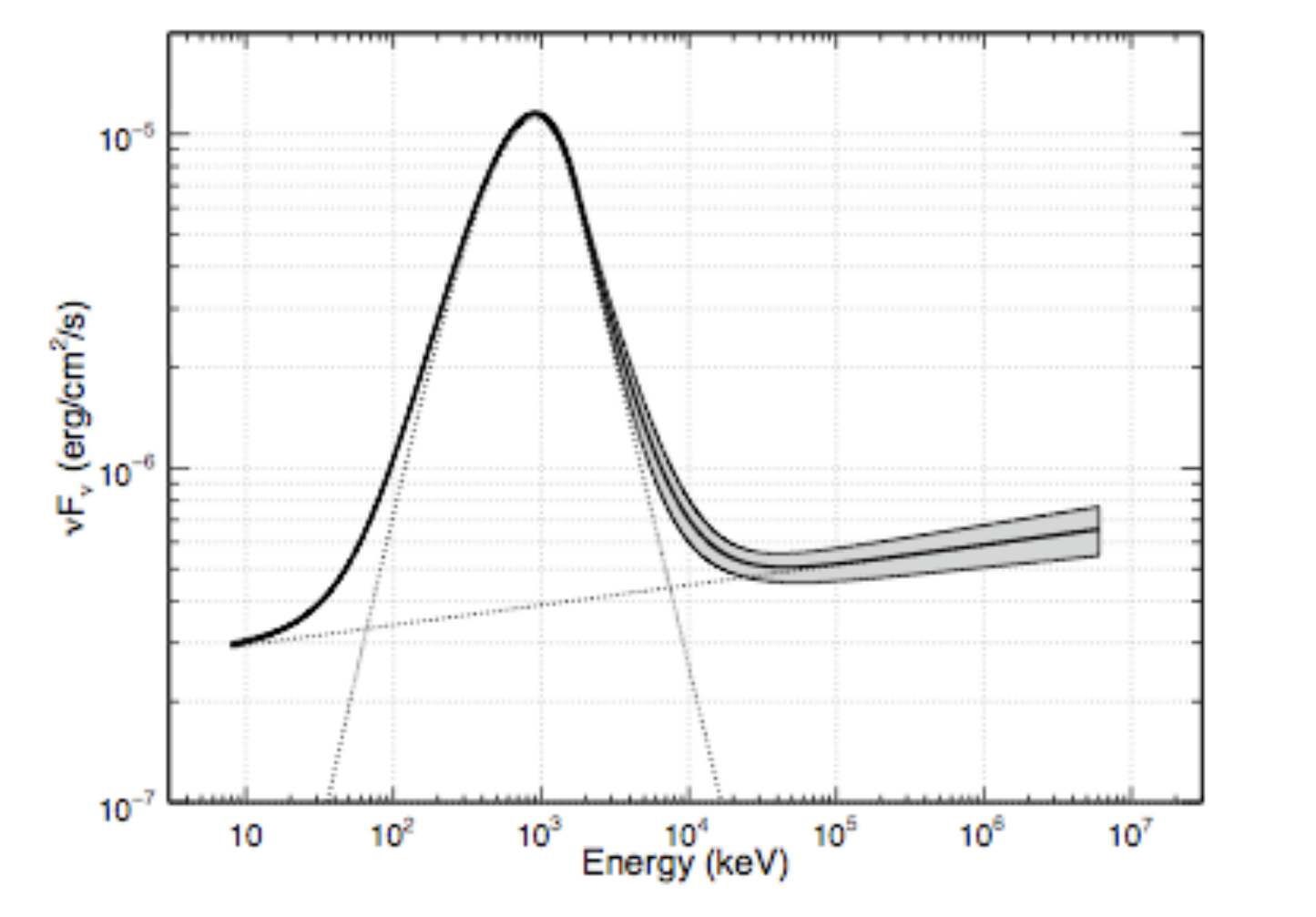}
  \caption{Top: The fluence spectrum of GRB 100724B, showing a deviation from the Band function
that is well fit by a thermal component \cite{guiriec:100724}.
Bottom: The $\nu$F$_\nu$ spectral fit for GRB 090902B from GBM and LAT data.  Excesses above the Band function are
well-fit by a power-law component that is present at low and high energies. \cite{grb090902b}}
  \label{fig:090902b}
 \end{figure}

The map in Figure \ref{skyfig} also shows the synergy between GBM and the Swift satellite, with 64
GRBs triggered in common by GBM and the Swift Burst Alert Telescope (BAT).
Bursts observed by the Swift BAT are likely to be localized well enough for follow-up, first by the X-ray and UV instruments on
Swift, and then by sensitive optical telescopes that can both measure the redshift and detect the host galaxy of the GRB
progenitor.   The Swift BAT, unlike GBM, has limited energy coverage,
so that studying the energetics of the prompt gamma-ray emission is often difficult.  The combination of the
broad-band afterglow energetics and host characteristics of Swift-detected bursts, with the knowledge of
prompt gamma-ray emission spectroscopy provided by GBM,
yields a much clearer picture for the bursts detected in common.  Studies of such bursts
include GRB 080810 \cite{page080810} and GRB 091024 \cite{gruber091024},
and ensemble analyses of the common GBM-Swift sample are presented in
\cite{racusin} for the afterglow, \cite{virgili} for the prompt emission, and \cite{gruber:rest} who estabish
the rest-frame properties of a sample of GBM-detected bursts which have redshift measurements.

GRB spectra have typically been fit using the Band function \cite{band}, two power laws smoothly joined with curvature
depending on the power law indices and a characteristic energy defined as the energy at which peak power per energy decade
($\nu$F$_\nu$) occurs.  The Band function is an empirical parameterization of
what is believed to be non-thermal radiation from accelerated particles in a relativistic outflow,
most likely synchrotron emission.
 The GBM spectral catalog \cite{ag:cat} confirms the preference for the Band function, or the closely related
smoothly-broken power-law function, over simpler models such as power-laws, or power-laws with exponential cut-offs,
in most of the bursts where the photon statistics allow parameters to be constrained despite the complexity of the fit.
In some bright bursts, however, the GBM data show the presence of deviations from the Band function at both low and
high energies, indicating that this empirical function may not always represent well the physical processes in GRBs.
Notably, GRB100724B exhibits a spectral bump that is well fit by a thermal component underlying the
non-thermal emission (Figure \ref{fig:090902b}, top).
The relative magnitudes of the thermal (fit by a blackbody function) and non-thermal (fit
by the Band function) components are used
to deduce that the relativistic outflow must be magnetized \cite{guiriec:100724}.
GRB090902B \cite{grb090902b} shows a large excess at low energies in the GBM data and
the deviation from the Band function is obvious over a wide energy range.  At higher energies, the LAT also sees
this deviation as a high-energy excess above the Band function.   A representation of
the overall spectral shape of GRB090902B in $\nu$F$_\nu$ is shown in the lower panel of Figure \ref{fig:090902b}, with the extra
spectral component visible both at low energies in GBM and at high energies in the LAT.  In fact, this behavior is common
among LAT-detected GRBs, with all but one \cite{grb080916c}
 of the bright LAT-detected bursts showing evidence in the LAT data for excesses
above the Band function extrapolation from GBM to high energies \cite{grb090926a, grb090510}.
It is difficult to associate the extra high-energy spectral component with a particular physical mechanism as it does not exhibit
the spectral turnover one might expect for an inverse Compton contribution, although statistics are poor in the multi-GeV
regime for LAT GRBs and such a turnover might occur at energies above this.  More seriously, one
would not expect an inverse Compton component to extend to low energies and account for the GBM low-energy excess.
If the high-energy excess is indeed an inverse Compton bump it is
likely these deviations from the Band function are not explained by a common physical mechanism across six
decades of energy, but do point to the inadequacy of the Band function to represent the GRB physical
emission mechanisms over a broad energy range.
The wealth of spectral
data provided by {\it Fermi} has produced interpretations and modeling of GRB spectra beyond the Band function
including photospheric models \cite{felix},
and purely physical modeling of the emission as synchrotron radiation from
power-law distributions of electrons, without using empirical fits \cite{jmb}.

\begin{figure}
  \vspace{5mm}
  \centering
  \includegraphics[width=3.in]{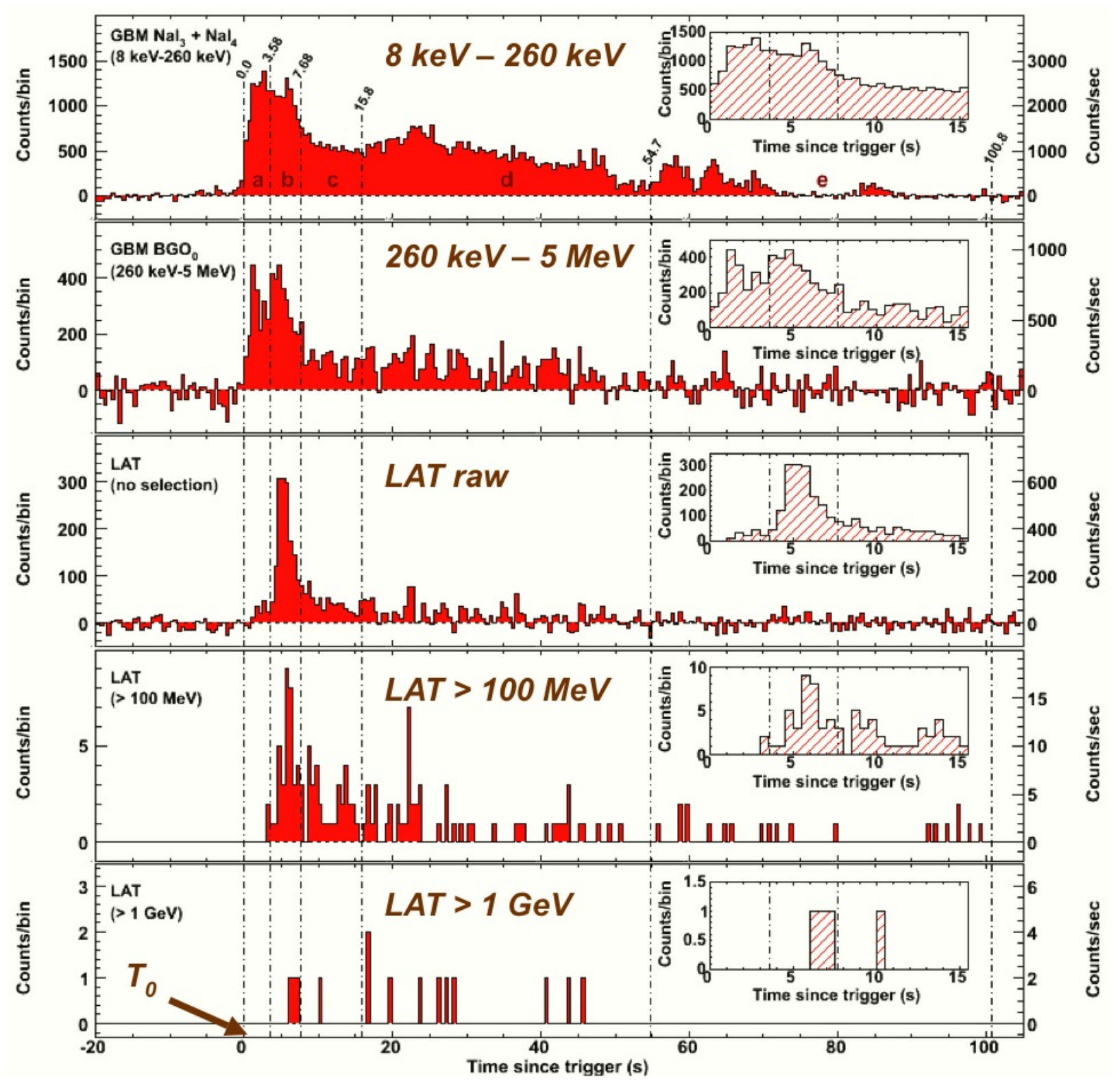}
  \caption{Lightcurve from GRB 080916C from low to high energies as seen by the GBM and LAT detectors.
Times are relative to the GBM trigger time.  It is seen that the first peak detected by GBM is not seen above 100 MeV.
The insets show a zoomed in picture of the lightcurve near the GBM trigger time. \cite{grb080916c}}
  \label{fig:080916c}
 \end{figure}

The ability to measure the prompt gamma-ray emission over such a wide energy band has yielded other surprises: in bright
LAT-detected bursts, it is clear that the onset of the high-energy emission lags that of the emission seen in GBM.
Figure \ref{fig:080916c} shows the lightcurve from GRB 080916C in multiple energy bands.  The first peak is missing
above 100 MeV, although joint GBM-LAT spectral fits do not suggest a cut-off is required, indicating that GRB spectra may evolve from
soft to hard before exhibiting the typical softening behavior over time observed by previous experiments.  Other surprises
include the persistence of the high-energy emission long after GBM ceases to discern impulsive peaks, although the
smoothness of this long-lived emission would make it difficult to detect in the background-limited GBM detectors.  Joint
Swift-LAT observations of GRB090510 \cite{mdp:090510} suggest this long-lived LAT emission is more akin to afterglow
radiation than continued central engine activity.  Finally, the emergence of intense radiation from very small source regions
(inferred by the variability time-scales of GRB peaks) with no evidence for quashing of the high-energy emission
through pair-production interactions with the dense lower-energy photon fields, implies very large mimimum Bulk Lorentz factors
in the relativistic outflows.  These factors could be as large as 1000, though a
geometry with multiple emission zones reduces this minimum Lorentz value by a factor of about 5 \cite{zou}.

\begin{figure}
  \vspace{5mm}
  \centering
  \includegraphics[width=3.4in]{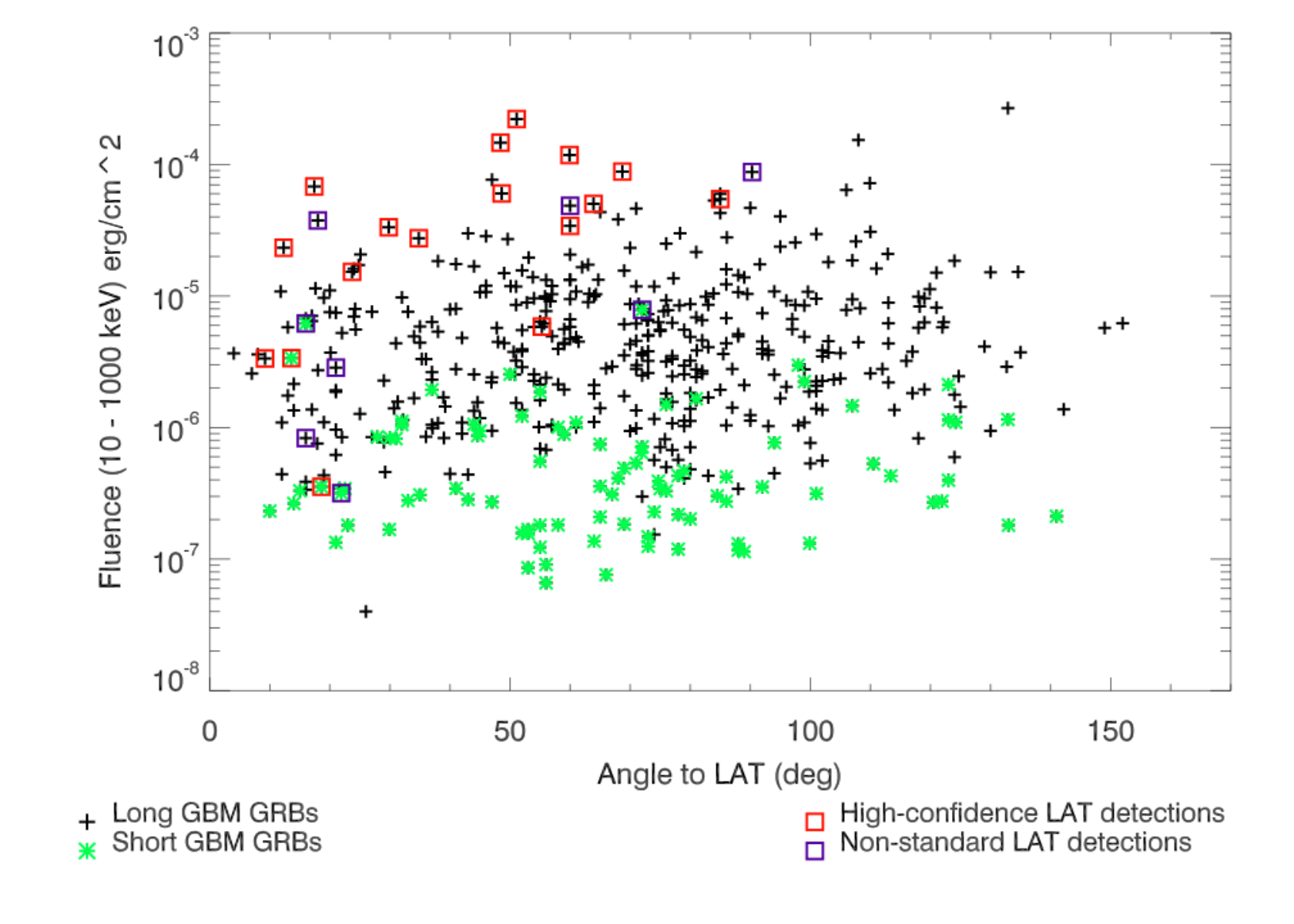}
  \caption{Fluence (10 - 1000 keV) measured by GBM for nearly 500 GRBs in the GBM spectral catalog \cite{ag:cat}
as a function of angle from the LAT boresight.  The short bursts are light asterisks, the long bursts are dark pluses.
Open squares indicate the LAT-detected bursts, which are among the brightest bursts seen by GBM.  The LAT sensitivity
to dimmer GBM bursts improves at low angles to the boresight.}
\label{fig:flu}
\end{figure}

Generalizations about high-energy emission from GRBs are difficult given the small number of bright GRBs detected by the LAT.
Figure \ref{fig:flu} shows the range of fluences seen in the GBM catalog bursts, plotted against angle to the LAT boresight,
with the color-coding distinguishing short from long bursts, and open squares showing the LAT-detected GRBs.
Although the fluence range sampled by GBM is broad,
the LAT is seeing only the brighter short and long GRBs seen by GBM, with greater
sensitivity to weaker bursts on-axis and when detection criteria are loosened.
To first order the detectability at high energies is determined by the brightness at low energies
rather than the low-energy spectral parameters, which is
as one might expect if the LAT is seeing extra spectral components that
are not clear extrapolations of the low-energy spectral form.

\section{Galactic Sources}

\subsection{Soft Gamma-ray Repeaters}
The main galactic (non-solar system) source
of GBM triggers is the population of highly-magnetized neutron stars known as magnetars, which includes Soft Gamma-ray Repeaters
(SGRs) and Anomalous X-ray Pulsars (AXPs).
Bursts from magnetars are short ($<$ 100 ms),
spectrally softer than GRBs, and can occur in isolation or as multiple event episodes during an
SGR activation.  As shown in Figure \ref{trigs}, GBM has seen several activations from magnetars, beginning the month after
launch, with two sources responsible for the three main outbursts.  No large outbursts have been seen since April 2009, although
three additional magnetars have shown bursting activity detected by GBM, including SGR 0418+5279, which was discovered
by GBM and
by other instruments which detected at least one of the three bursts seen by GBM from this source \cite{newsgr}.

\begin{figure}
  \vspace{5mm}
  \centering
  \includegraphics[width=3.in]{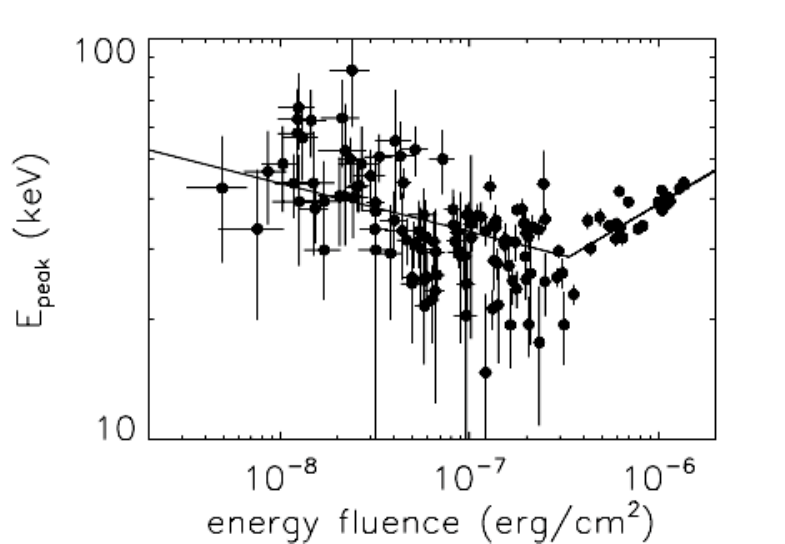}
  \caption{Energy fluence versus peak energy for time-resolved spectral fits to the 5 brightest bursts seen by
   GBM from SGR 0501+4516 in its 2008 outburst. \cite{lin}}
  \label{lin:epeak}
 \end{figure}

Spectral and temporal properties from these outbursts have been published \cite{kaneko, lin, avdh}.
One of the significant discoveries of Lin et al. \cite{lin},
who analyzed the outburst of SGR 0501+4516, utilized the unique broad band spectral capability of GBM to settle an
 open issue in the field. The energy spectrum of SGR bursts is often represented either by a thermal bremsstrahlung model or
by the so-called Comptonized model, a power-law which decays exponentially above a peak energy.
Previous experiments had measured an inverse relationship between the intensity and the hardness of SGR
 bursts \cite{sgr:ersin}, whilst others saw a direct relationship \cite{sgr:fenimore}.
  The earlier studies used data from two distinct SGRs, which were different again from the source studied in \cite{lin}.
The fluence range over which GBM detects bursts encompasses
those of the samples used in both of these studies owing to the combination of a spectral range well-suited
 for magnetar spectroscopy, wide field-of-view, duty cycle, and sensitivity.
 Lin et al \cite{lin} show that both observations are correct: Figure \ref{lin:epeak} shows that there is a
minimum fluence above which the direct relationship between hardness (expressed for the first time as peak energy) and burst
fluence holds, whilst below this fluence the inverse relationship is seen.

 It is difficult to see the quiescent emission from magnetars with GBM, but searches for periodicity
 in the GBM continuous data between 27 and 300 keV
  show the recovery of the neutron star period for four magnetars \cite{kuiper}.

\subsection{X-Ray Binary Systems}
\begin{figure}
  \vspace{5mm}
  \centering
  \includegraphics[width=3.in]{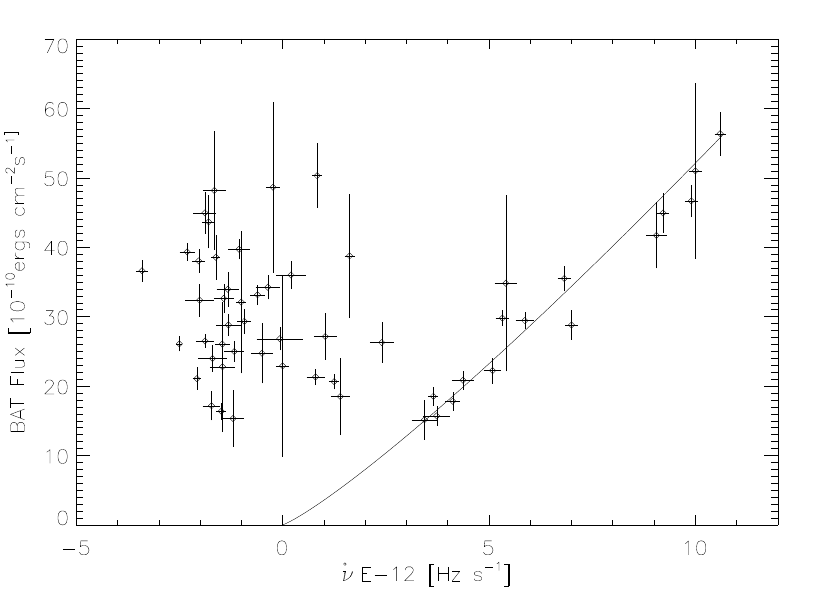}
  \caption{Spin-up rate for the X-ray binary system OAO 1657-415 as a function of hard X-ray flux.  At high spin-up rates
a correlation is clear, implying the presence of an accretion disk.  In other states, there is no correlation. \cite{jenke}}
  \label{torque}
 \end{figure}

Accreting binary systems are prime candidates for both the Earth Occultation and pulsar techniques using GBM data.  To date,
86 such systems have been detected using Earth Occultation (of which
74 are neutron-star binaries), and 26 neutron-star systems are seen via their pulsed emission, whether persistent or in outburst.
The value of GBM observations of these sources comes from the continuous monitoring of their flux, spectrum,
and frequency, which can alert narrow-field instruments as well as observers at other wavelengths that the source is undergoing
an outburst or a state transition, frequency glitch, or even a torque reversal. Optical monitoring of
the binary system A 0535+26 was correlated with GBM-measured outbursts of the source yielding broadband measurements
of the source during a giant outburst in which QPOs were seen in the GBM data \cite{choni:a0535}. Torque switching is
seen in several binary systems monitored by GBM.
The High-Mass X-ray Binary OAO 1657-415 shows a correlation between flux level and
spin-up frequency at high spin-up rates, indicating that an accretion disk has formed in this wind-driven system (Figure
 \ref{torque}).  At
lower spin-up rates and when the star is spinning down, there is no such relationship, despite the low-spin
flux levels spanning
an intensity range comparable to that in the high spin-up state \cite{jenke}.  It is not known
 what causes this accretion-disk to
form and persist, and overall the source's long-term behavior is interesting, with an orbit that appears to be decaying over
time.  Other systems exhibit rarer torque reversals, including the
ultra-compact Low-Mass X-ray Binary 4U 1626-67, which before the launch of {\it Fermi}
had last been observed by Chandra in a spin-down state. GBM found the source in a spin-up state, its first
torque reversal since the Ginga-ROSAT era \cite{choni}.

Some low-mass X-ray binary systems are also known for their bursting activity, with Type 1 X-Ray bursts indicating thermonuclear
explosions on the accreting neutron star.
GBM is sensitive to these events, although their peak shapes (long and smoothly rising)
and their soft thermal (temperatures of 2.5 - 3 keV)
spectra make it unlikely
that such bursts will trigger the instrument.  A search for intermediate and long
Type 1 X-Ray Bursts in the GBM continuous data
has uncovered many candidates, particularly for long Type 1 bursts. GBM
is expected to see only the tip of the emission, and with peaks lasting tens of seconds in GBM, we
expect these are bursts lasting
15 minutes or so in more sensitive X-ray detectors.
Whilst the uncollimated nature of the GBM detectors
means source confusion is a problem in crowded regions, the collection of long bursts clearly associated with 4U 0614+09
has allowed us to place recurrence time-scale estimates on this source (in
a low mass-accretion state) of $16 \pm 6$ days \cite{linares}.

Black hole binary systems have also proved revealing, with long-term Earth Occultation monitoring
of flux levels over a broad energy range
from Cyg X-1
pointing to five state transitions that
were simultaneously or subsequently widely observed by other satellites and ground-based
telescopes \cite{case:cyg}.
Black hole systems are among the high-confidence Earth Occultation detections above 100 keV, making GBM particularly useful
to monitor these sources for state transitions indicated by their energy-dependent flux variations.

\subsection{The Crab Nebula}
\begin{figure}
  \vspace{5mm}
  \centering
  \includegraphics[width=3.2in]{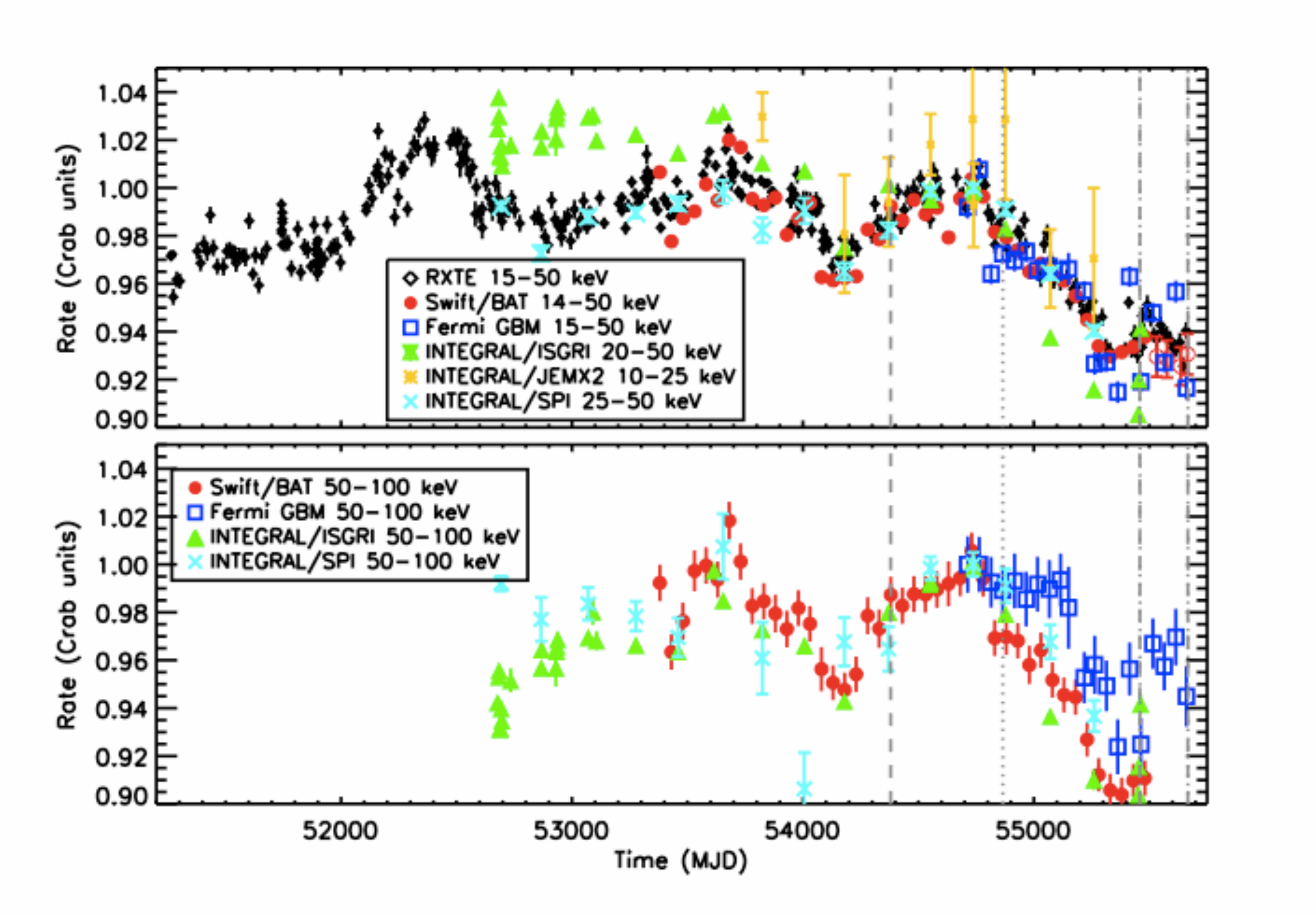}
  \caption{Composite Crab light curves for RXTE/PCA (top only - black diamonds),
Swift/BAT (top \& bottom, red circles), Fermi/GBM (top \& bottom - blue squares), INTEGRAL/ISGRI (top \& bottom, green triangles), INTEGRAL/JEM-X2 (top only, orange asterisks), and INTEGRAL/SPI (top \& bottom - light blue crosses). Each data set has been normalized to its mean rate in the time in- terval MJD 54690-54790. Vertical lines denote GeV flares observed with AGILE (dotted) and Fermi/LAT (dashed). \cite{cwh:crab}}
  \label{crab}
 \end{figure}

One of the brightest sources uncovered in the Earth Occultation measurements has proved unexpectedly rich.
The Crab Nebula, long used as a standard candle for X-ray astronomy, showed a decline in flux of 7\% over the first
two years of GBM Earth Occultation measurements.  After verification using observations by other instruments, it has emerged that
the Crab is actually a variable source over a long baseline in time \cite{cwh:crab}, as can be seen in Figure \ref{crab}.

\section{Solar System Science}
\subsection{The Sun}
\begin{figure}
  \vspace{5mm}
  \centering
  \includegraphics[width=3.in]{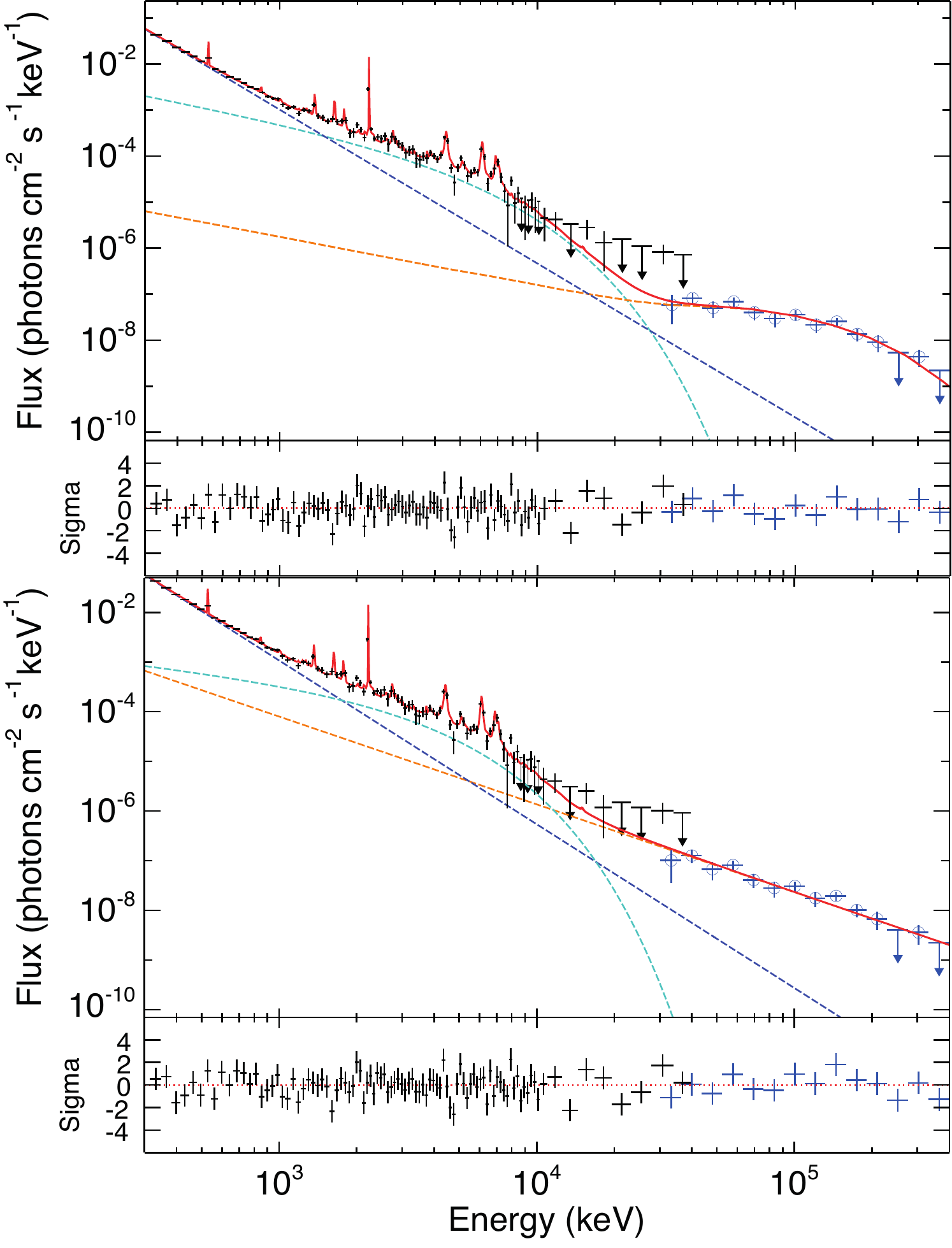}
  \caption{Spectral fits to the GBM and LAT data from the M-class flare seen on June 12 2010. At low energies,
two distinct non-thermal components are needed, in addition to the nuclear, annihilation, and neutron capture lines seen
in the GBM BGO data.  At higher energies, the emission can be fit either as bremsstrahlung
 from higher-energy electrons (top) or gamma rays from pion decay resulting from accelerated high-energy protons (bottom).
\cite{solar}}
  \label{fig:solar}
 \end{figure}

Solar flare triggers have become increasingly common as this solar cycle reaches its peak.  On June 12 2010, GBM registered
its first flare with measurable nuclear lines, which were measured up to a few
MeV, in addition to the neutron capture and annihilation lines.
The LAT saw photons up to 400 MeV. Although the joint GBM-LAT spectrum did
not distinguish between a model where the high-energy emission was attributed to bremsstrahlung-radiating electrons with energies
of hundreds MeV from one where hadronic primaries produced pions that decayed into gamma rays, several non-thermal
components were required to explain the broad-band emission underlying the nuclear lines. The intensity of the line emission
was quite surprising for an M-class flare, and the most interesting aspect of this flare is that the MeV bremsstrahlung,
the high-energy emission, and the line emission were temporally coincident within seconds, implying that acceleration
to 100s MeV from the mildly relativistic electrons responsible for the lower-energy bremsstrahlung component
took only seconds \cite{solar}.
Figure \ref{fig:solar}  shows the spectral fit to this flare, with the two different fits to
the high-energy component featured in separate panels.

Searches for quasi-periodic pulsations (QPPs) from bright GBM-detected solar flares revealed no such behavior.
The authors in \cite{gruber:solar}
exposed possible methodological problems with previous searches that had uncovered QPPs in flares detected
by the Reuven Ramaty High Energy Solar Spectroscopic Imager (RHESSI \cite{rhessi:lin}).

\subsection{Terrestrial Gamma-ray Flashes}

\begin{figure*}
  \vspace{5mm}
  \centering
  \includegraphics[width=6.in]{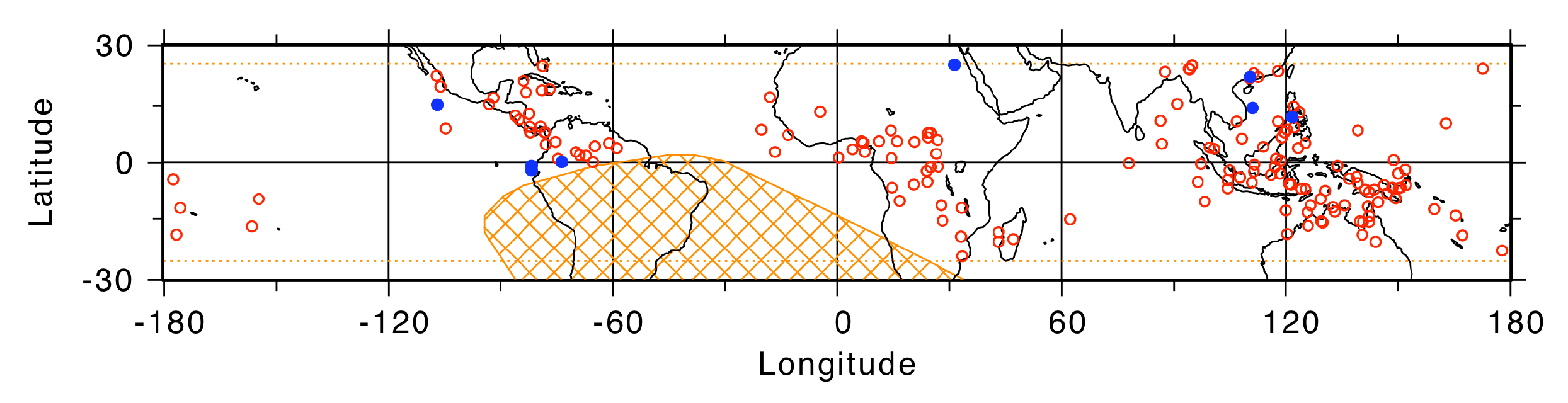}
  \caption{The position of {\it Fermi} above the Earth at the time of TGF triggers.  Open circles indicate
gamma-ray TGFs and follow regions of thunderstorm activity. Closed circles show the position of {\it Fermi} at the time
of electron TGF triggers, along the magnetic field lines joining Fermi to the magnetic footprint where the TGF originated.}
  \label{fig:subsc}
 \end{figure*}

In the first three years of operation, GBM triggered on 170 Terrestrial Gamma-ray Flashes (TGFs), with the rate increasing
from one per month to two per week following flight software changes in November 2009.  TGFs are brief
(usually shorter than 1 ms) hard bursts associated with electron
acceleration in the electric fields near the tops of thunderstorm clouds where lightning activity occurs.  Fermi has
a detection horizon for TGFs of several hundred kilometers, so that if one looks at the sub-spacecraft position at the times of TGF
triggers (Figure \ref{fig:subsc}) the cumulative pattern follows the distribution of lightning activity within the {\it Fermi}
orbital boundaries.  Coincident VLF radio signals measured with the World Wide Lightning Network (WWLLN) allow the geolocation
of 30\% of GBM-detected TGFs, and strengthen the association of TGFs with particular regions of storm activity \cite{vc}.
The discharges measured by WWLLN are typically simultaneous with the TGF emission.
Closer relationships between TGFs and radio signals can be established using instrumentation that records the
discharge waveform.  The Duke University detectors were used to show that the radio waveform closely follows the
gamma-ray pulse shape of the TGF \cite{cummer}.
Temporal characteristics of GBM TGFs are presented in \cite{briggs1, fishman}, showing that GBM is sensitive
to single, multiple and overlapping TGF peaks, with pulse fitting of 150 TGFs indicating
two distinct pulse shapes, one of them asymmetric and one Gaussian \cite{foley}.

\begin{figure*}
  \vspace{5mm}
  \centering
  \includegraphics[width=3.5in]{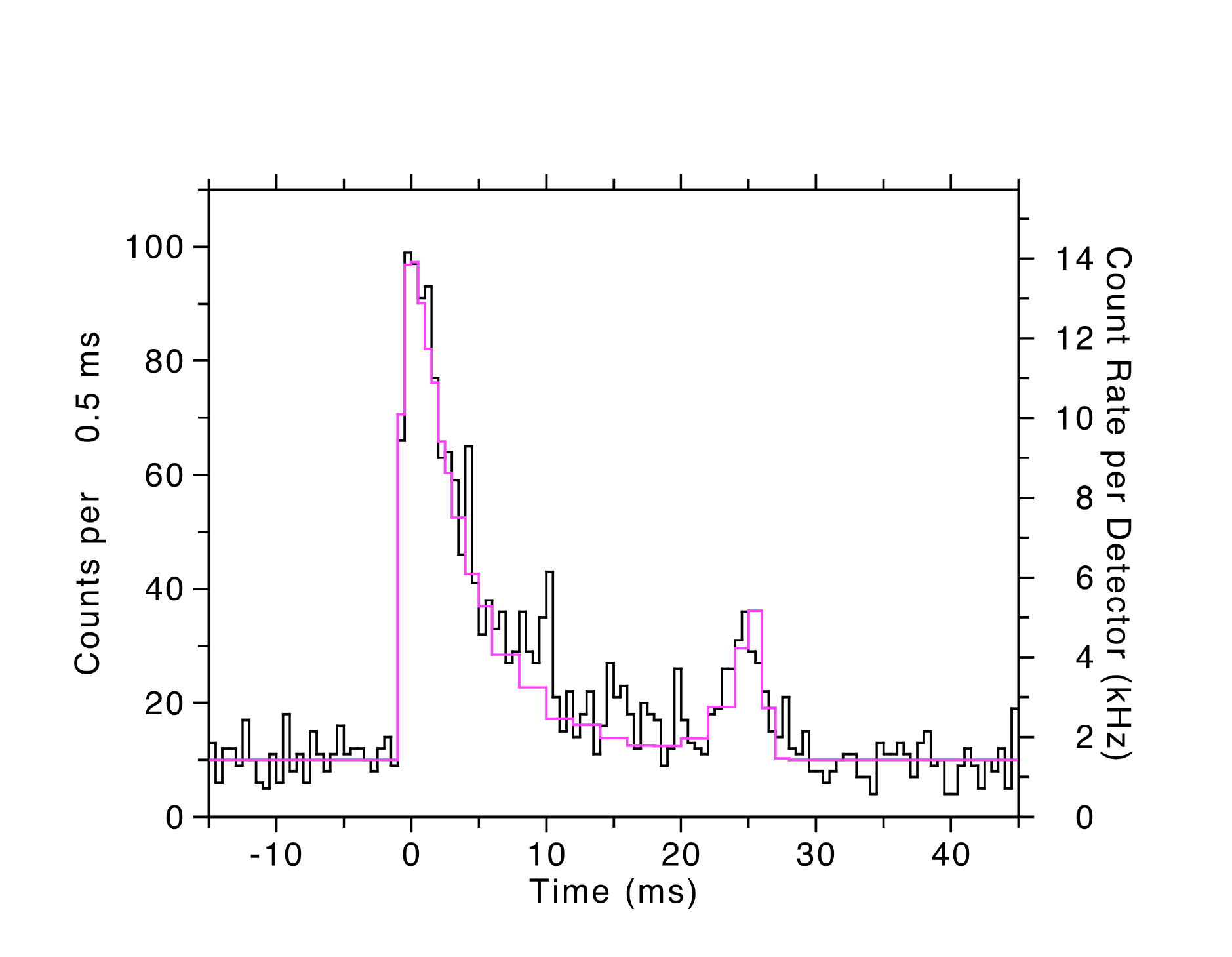}
  \includegraphics[width=2.5in]{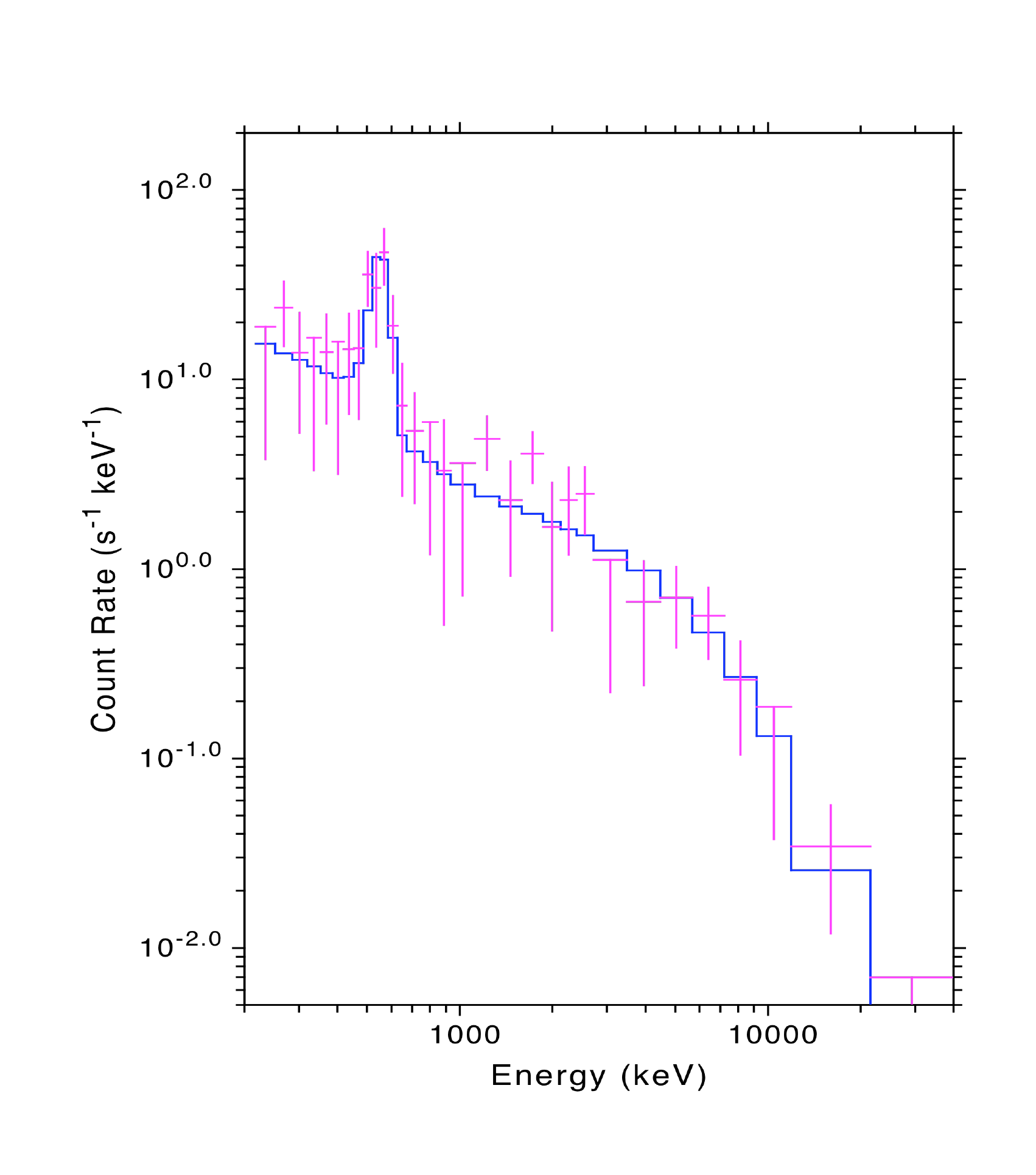}
  \caption{Electron TGFs have characteristic signatures.  The left panel shows the lightcurve from TGF 091214 which originated
thousands of km from the {\it Fermi} nadir.  Electrons traveled along a magnetic field line joining the originating
thunderstorm to {\it Fermi}, some were detected as a strong peak on impact with GBM detectors, others
underwent magnetic mirroring high enough in the
atmosphere to survive extinction and double back to {\it Fermi}, making the second, smaller peak. The right panel shows the energy
spectrum from this electron TGF.  The continuum is a good fit to an exponential distribution of electrons, but the bump
indicates a large (20\%) fraction of positrons which annihilate on impact with material in {\it Fermi} to give a gamma-ray
line at 511 keV. \cite{briggs2}}
  \label{fig:electron}
 \end{figure*}

A smaller horizon applies to those TGFs which are detected as an electron beam rather than as bremsstrahlung emission from
the electron cascade.  These electron TGFs are much rarer, occurring when electrons in the cascade are captured by a
magnetic field line, spiralling along the field line with a distribution of pitch angles and path lengths,
and captured by {\it Fermi} which is positioned along the magnetic field line and can be thousands of km away.  Electron beam TGFs
are longer than gamma-ray TGFs and were probably detected in both BATSE and RHESSI instruments.
Measurement of radio signals from
lightning at one of the geomagnetic footprints for one electron TGF \cite{cohen} seen by GBM  strengthened the case.
Magnetic mirroring of the electrons from
another electron beam TGF seen by GBM produces a secondary peak with the expected peak
separation for the field line geometry and the position of {\it Fermi} (Figure \ref{fig:electron}, left).
The decisive measurement, however, was the detection of strong annihilation lines at 511 keV in the GBM
spectra for bright electron TGFs (Figure \ref{fig:electron}, right)
indicating the presence of positrons which annihilated to produce gamma-ray lines upon
impact with {\it Fermi}.  The strength of the lines indicates that the electron beams have an antimatter fraction of about 20\%
 \cite{briggs2}.

\section{Conclusions}
In its first three years of operation, GBM has enabled a broad scientific program covering sources
as close as 10 km above the Earth's surface out to explosions in the earliest universe.
The instrument has functioned perfectly from the day the high voltages were turned on in June 2008, and continues to operate
without problems.  Improvements by the instrument team in the areas of source localization, detector calibration, and new
operating modes that will provide high temporal resolution data over much or all of the orbit, are in progress.  All GBM
data are publicly available at the FSSC\footnote{
http://fermi.gsfc.nasa.gov/ssc/data/access/gbm/}.

\clearpage


\vspace{2in}

\begin{thebibliography}{}


\bibitem{grb080916c} Abdo, A. et. al., Science, 2009, {\bf 323}, 1688.
\bibitem{grb090902b} Abdo, A. et. al., ApJL, 2009, {\bf 706}, L138.
\bibitem{grb090510} Ackermann, M. et. al., ApJ, 2010, {\bf 716}, 1178.
\bibitem{grb090926a} Ackermann, M. et. al., ApJ, 2011, {\bf 729}, 114.
\bibitem{solar} Ackermann, M. et. al., ApJ, 2012, {\bf 745}, 144.
\bibitem{atwood} Atwood, W.B. et al., ApJ, 2009, {\bf 697}, 1071.
\bibitem{band} Band, D.L. et al., ApJ, 1993, {\bf 413}, 281.
\bibitem{bhat:pulses} Bhat, P.N. et al., ApJ, 2012, {\bf 744}, 141.
\bibitem{eb} Bissaldi, E. et al., ApJ, 2011, {\bf 733}, 97.
\bibitem{briggs1} Briggs, M.S. et al., J.Geophys.Res., 2010, {\bf 115}, A07323.
\bibitem{briggs2} Briggs, M.S. et al., Geophys.Res.Lett., 2011, {\bf 38}, L02808,5.
\bibitem{jmb} Burgess, J.M. et al., ApJ, 2011, {\bf 741}, 24.
\bibitem{choni:a0535} Camero-Arranz, A. et al., submitted to ApJ.
\bibitem{choni} Camero-Arranz, A. et al., ApJ, 2010, {\bf 708}, 1500.
\bibitem{case} Case, G.L. et al., ApJ, 2011, {\bf 729}, 105.yy
\bibitem{case:cyg} Case, G.L. et al., submitted to ApJ.
\bibitem{cohen} Cohen, M.B. et al., Geophys.Res.Lett., 2010, {\bf 37}, L18806,4.
\bibitem{vc} Connaughton, V. et al., J.Geophys.Res., 2010, {\bf 115}, A12307.
\bibitem{vc:loc} Connaughton, V. et al., in preparation.
\bibitem{corsi} Corsi, A. et al., ApJ, 2011, {\bf 741}, 76.
\bibitem{cummer} Cummer, S. et al., Geophys.Res.Lett., 2011, {\bf 38}, L14810.
\bibitem{sgr:fenimore} Fenimore, E.E. et al., ApJ, 1994, {\bf 432}, 742.
\bibitem{fishman} Fishman, G.J. et al., J.Geophys.Res., 2011, {\bf 116}, A07304.
\bibitem{foley} Foley, S. et al., submitted to J.Geophys.Res.
\bibitem{ghirlanda} Ghirlanda, G., Ghisellini, G., \& Nava, L., MNRAS, 2011, {\bf 418}, 109.
\bibitem{sgr:ersin} G\"{o}\v{g}\"{u}\c{s}, E. et al., ApJ, 2001, {\bf 558}, 228.
\bibitem{ag:cat} Goldstein, A., to appear in ApJSupp.
\bibitem{gruber:rest} Gruber, D. et al., A\&A, 2011, {\bf 531}, 20.
\bibitem{gruber091024} Gruber, D. et al., A\&A, 2011, {\bf 528}, 15.
\bibitem{gruber:solar} Gruber, D. et al., A\&A, 2011, {\bf 533}, 61.
\bibitem{guiriec:short} Guiriec, S. et al., ApJ, 2010, {\bf 725}, 225.
\bibitem{guiriec:100724} Guiriec, S. et al., ApJ, 2011, {\bf 727}, 33.
\bibitem{newsgr} van der Horst, A. et. al., ApJL, 2010, {\bf 701}, L1.
\bibitem{avdh} van der Horst, A. et. al., to appear in ApJ.
\bibitem{jenke} Jenke, P. et al., submitted to ApJ.
\bibitem{kaneko} Kaneko, Y. et al.,  ApJ, 2010, {\bf 710}, 1335.
\bibitem{kuiper} Kuiper, L., ter Beek, F., \& Hermsen, W., in preparation for submission to ApJ.
\bibitem{lin} Lin, L. et. al., ApJ, 2011, {\bf 739}, 87.
\bibitem{rhessi:lin} Lin, R. et al., Sol. Phys., 2002, {\bf 210}, 3.
\bibitem{linares} Linares, M. et al., 2011, Proc. 4th MAXI workshop, http://maxi.riken.jp/FirstYear, and in preparation.
\bibitem{meegan09} Meegan, C.A. et. al., ApJ, 2012, {\bf 702}, 791.
\bibitem{page080810} Page, K.L. et al. MNRAS, 2009, {\bf 400}, 134.
\bibitem{wsp:cat} Paciesas, W.S. et al., to appear in ApJSupp.
\bibitem{mdp:090510} De Pasquale, M. et. al., ApJL, 2010, {\bf 709}, L146.
\bibitem{racusin} Racusin, J. et al., ApJ, 2011, {\bf 738}, 138.
\bibitem{felix} Ryde, F. et. al., ApJL, 2010, {\bf 709}, L172.
\bibitem{soderberg} Soderberg, A.M., Nature, 2010, {\bf 463}, 513.
\bibitem{virgili} Virgili, F. et al. submitted to MNRAS.
\bibitem{cwh:crab} Wilson-Hodge, C.A. et al., ApJL, 2011, {\bf 727}, L40.
\bibitem{cwh:catalog} Wilson-Hodge, C.A. et al., submitted to ApJSupp.
\bibitem{zou} Zou, Y., Fan, Y., \& Piran, T., ApJ, 2011, {\bf 762}, 2.

\end{thebibliography}
\end{document}